\documentclass[nofootinbib,notitlepage,superscriptaddress,10pt,aps,prd,twocolumn]{revtex4-1}

\usepackage[utf8]{inputenc}
\usepackage{amsmath,mathtools}

\usepackage{tensor}
\usepackage{amssymb}
\usepackage{graphicx}
\usepackage{physics}
\usepackage{lipsum}  
\usepackage{bigints}
\usepackage{verbatim}
\newcommand{\leri}[1]{\left(#1\right)}

\usepackage{braket}
\usepackage{cancel}
\usepackage{color}
\usepackage{hyperref}

\newcommand{\intgr}{\frac{1}{16\pi}\int d^4x \sqrt{-g}}
\newcommand{\intgrE}{\frac{1}{16\pi}\int d^4x \sqrt{-\tilde{g}}}

\begin{document}
\title{Super-entropic black hole with Immirzi hair}
\author{Simon Boudet}
\email{simon.boudet@unitn.it}
\affiliation{Dipartimento di Fisica, Universit\`{a} di Trento,\\Via Sommarive 14, I-38123 Povo (TN), Italy}
\affiliation{Trento Institute for Fundamental Physics and Applications (TIFPA)-INFN,\\Via Sommarive 14, I-38123 Povo (TN), Italy}
\author{Flavio Bombacigno}
\email{flavio.bombacigno@ext.uv.es}
\affiliation{Departament de F\'{i}sica Teòrica and IFIC, Centro Mixto Universitat de València - CSIC, Universitat de València, Burjassot 46100, València, Spain}
\author{Giovanni Montani}
\email{giovanni.montani@enea.it}
\affiliation{Physics Department, ``Sapienza'' University of Rome,\\ P.le Aldo Moro 5, 00185 (Roma), Italy}
\affiliation{ENEA, Fusion and Nuclear Safety Department,\\ C. R. Frascati,
 	Via E. Fermi 45, 00044 Frascati (Roma), Italy}
\author{Massimiliano Rinaldi}
\email{massimiliano.rinaldi@unitn.it}
\affiliation{Dipartimento di Fisica, Universit\`{a} di Trento,\\Via Sommarive 14, I-38123 Povo (TN), Italy}
\affiliation{Trento Institute for Fundamental Physics and Applications (TIFPA)-INFN,\\Via Sommarive 14, I-38123 Povo (TN), Italy}

\begin{abstract}In the context of $f(R)$ generalizations to the Holst action, endowed with a dynamical Immirzi field, we derive an analytic solution describing asymptotically Anti-de Sitter black holes with hyperbolic horizon. These exhibit a scalar hair of the second kind, which ultimately depends on the Immirzi field radial behaviour. In particular, we show how the Immirzi field modifies the usual entropy law associated to the black hole. We also verify that the Immirzi field boils down to a constant value in the asymptotic region, thus restoring the standard Loop Quantum Gravity picture. We finally prove the violation of the reverse isoperimetric inequality, resulting in the super-entropic nature of the black hole, and we discuss in detail the thermodynamic stability of the solution.
\end{abstract}

\maketitle

\section{Introduction}
A consistent quantum description of the gravitational interaction is maybe one of the most prominent challenges in modern theoretical physics. Among candidate theories of quantum gravity, Loop Quantum Gravity (LQG) constitutes an intriguing attempt to pursue a non perturbative, canonical quantization of General Relativity (GR) \cite{Rovelli2004,Thiemann2007}. The theory can be formulated at the Lagrangian level by including the Holst term \cite{Holst1996} into the Palatini version of the GR action, where the metric and the connection are treated a priori as independent variables (first order formalism). This additional term is driven by the so called Immirzi parameter \cite{Immirzi:1996di,Immirzi1997,Rovelli:1997na}, which turns out to play a fundamental part in constructing a viable gauge $SU(2)$ representation of the theory, by means of the Ashtekar variables \cite{Ashtekar1986,Ashtekar1987,Ashtekar1989,Ashtekar1992}. Here, we do not deal with the issues concerning its role in the quantizing procedure, and we rather follow the idea in \cite{Taveras2008,Calcagni2009,Bombacigno2016}, where it is promoted to a dynamical scalar field with the aim of investigating its properties. In particular, we decided to adopt a modified gravity perspective as in \cite{Veraguth2017,Wang2018,Wang2020,Bombacigno:2019nua,Iosifidis:2020dck}, by considering a Palatini $f(\mathcal{R})$-like generalization \cite{Olmo2011} of the Holst action in the presence of an Immirzi scalar field.
The resulting theory is equivalent to a non-minimally coupled scalar-tensor theory with the scalar sector including both the Immirzi field and the scalar field that encodes the degree of freedom of the $f(\mathcal{R})$ gravity in the Jordan frame (often called the scalaron).
\\Two main features can be traced back to the first order formulation characterizing the model. On one hand, torsion is present in the theory \cite{SHAPIRO2002113}, acquiring a dynamical character from the scalar fields. This allows us to fully solve connections in terms of metric and scalar fields, i.e. to work with an effective metric action (second order formalism), obtained by solving the torsion components in terms of the gradients of the scalar fields via standard methods.
\\On the other hand, the scalaron is governed by the so called structural equation, as in Palatini $f(\mathcal{R})$ gravity \cite{Olmo2011}. However, while in standard Palatini $f(\mathcal{R})$ gravity this implies a constant scalar field in vacuum, in the case at hand it actually depends algebraically on the Immirzi field, which can in principle induce also non trivial behaviours.
\\The Immirzi field has already been investigated in cosmological models \cite{Taveras2008,Bombacigno2016,BombacignoFlavioandMontani2019}, in the presence of fermion fields \cite{Mercuri2006,Mercuri2008}, and in connection with the propagation of gravitational waves \cite{Bombacigno2019,Bombacigno2018}, revealing an interesting phenomenology, such as the existence of bouncing solutions and the presence of additional gravitational waves polarizations, together with implications at a more fundamental level regarding the strong CP problem \cite{PhysRevD.81.125015,PhysRevD.91.085017} and the chiral anomaly \cite{Mercuri2009a}. However, investigations on the vacuum spherically symmetric sector of models featuring an Immirzi field are scarce in literature (see the appendix of \cite{Torres-Gomez2009}), the main reason for this being the no-hair theorems \cite{Sotiriou_2015}. The latter state that spherically symmetric solutions in vacuum scalar-tensor theories are identical to those of GR, provided we make the crucial assumptions of asymptotic flatness and stationarity. In other words, these theorems prevent the existence of black hole solutions with a non-trivial radial profile for the scalar fields.
\\In spite of this, a growing number of hairy black hole solutions  have been found thanks to the fact that the no-hair theorem can always be evaded violating some of the hypothesis on which it stands (see e.g. \cite{Martinez2003,Martinez2004,PhysRevD.85.084035,Cisterna:2018mww}). Adopting this strategy, we are forced to take into account more involved solutions  than the stationary and asymptotically flat template.
In particular we will consider the case of asymptotically Anti-de Sitter (AdS) spacetimes. Although less prone to a direct astrophysical connotation with respect to the asymptotically flat or de Sitter solutions, AdS ones are of interest in the context of the AdS/CFT correspondence, especially in light of works oriented towards a connection between different approaches to quantum gravity \cite{Bazeia:2014xxa,PhysRevD.80.065003,Ellis:2011mz,Vaid2018}, in which the role of the Immirzi parameter (field) is taken into account. Besides, as it was firstly realized in \cite{Vanzo1997,Birmingham1999}, asymptotically AdS black holes allow for a wider variety of horizon topologies with respect to the usual spherical case. These topological black holes possess a horizon of constant curvature (positive, negative or vanishing), and they can form as the result of a gravitational collapse (see \cite{Smith1997}). We observe, moreover, that the Palatini reformulation usually enhances the appearance of non trivial structures in compact objects evolution, affecting their topology or the nature of the singularity \cite{Bambi:2015zch,Menchon:2017qed,Afonso:2019fzv,Olmo:2019flu}.
\\ \noindent In this paper we report an analytical solution describing an asymptotically Anti-de Sitter and topological black hole endowed with scalar hair provided by the Immirzi field. The solution reduces to the one found in \cite{Martinez2004} in certain limits of the model parameters.
The event horizon is a surface of constant negative curvature, i.e. it has a hyperbolic topology, which can be described as a compact surface of genus $g\geqslant 2$ via suitable identification of points on the hyperbolic plane \cite{Smith1997}.
\\The black hole hair are realized by both the 
Immirzi field and the scalaron, which interact via the modified structural equation. The presence of the black hole is able to excite the Immirzi field, which shows a non-trivial radial profile. In the large radius limit, the Immirzi field becomes a constant parameter and one recovers the standard LQG scenario.
\\Another interesting feature of black holes in AdS spacetimes is that they have a well defined thermodynamics, which we carefully analyse for our solution. Black hole thermodynamics has been a prolific research field since its first appearance \cite{Bekenstein1972,Hawking1975}. It is now well established the existence of laws of black hole thermodynamics describing these objects in terms of thermodynamic variables such as temperature and entropy (see \cite{doi:10.1142/S0218271814300237} for a review).
\\Although the physics of the microscopic degrees of freedom at the origin of such macroscopic properties is still not completely understood, it is likely rooted in some theory of quantum gravity. Thus, the semi-classical approach to black hole thermodynamics can shed light on the matter, yielding interesting clues \cite{Johnson2019}.
\\Since the first calculation performed by Hawking, several semi-classical methods have been developed  to derive thermodynamic quantities of interest. Among them, we mention the Euclidean path integral method \cite{PhysRevD.15.2752} and Wald's entropy formula \cite{PhysRevD.48.R3427}.
\\Here, we derive the thermodynamics of the black hole solution at hand by following the Euclidean path integral method. Particular attention has to be paid to the computation of the Euclidean action. Beside the appropriate Gibbons-Hawking-York (GHY) boundary term \cite{Dyer2009}, necessary to render the variational principle well-posed, a regularization procedure is needed to cure the divergence of the action. We take care of this by following the counter-terms method both in an implicit way, as in \cite{Martinez2004}, and by providing the explicit covariant expressions of the counter-terms, generalizing the ones illustrated in \cite{Myers1999,Gegenberg2003} to the case of a non minimally coupled scalar field.
\\ The expression for entropy obtained with this method shows that the Immirzi hair are responsible for a modification of the well-known area law $S=A/4$, which shows a correction due to the Immirzi field computed at the black hole event horizon.
We check our results by also applying Wald's formula, and the two procedures are consistent despite the presence of torsion in the theory \cite{CHAKRABORTY2018432}.
\\ Recently, it has been proposed a way to enlarge the thermodynamic phase space by including a pressure term, related to the cosmological constant, and its conjugate quantity, a thermodynamic volume. This extended phase space approach \cite{Kubiznak2017} has been widely examined revealing several analogies between the thermodynamics of black holes and the one of usual matter systems. In this context, the thermodynamic volume has initially been conjectured to satisfy the so called reverse isoperimetric inequality \cite{Dolan2013}, implying an upper bound on the amount of entropy a black hole can have at a given volume, the maximum being attained by the Schwarzschild-AdS black hole.
\\ However, there are several black hole solutions, dubbed super-entropic black holes, that violate the reverse isoperimetric inequality  \cite{PhysRevD.95.046002,PhysRevD.92.044015,PhysRevD.89.084007,PhysRevLett.115.031101,Feng2017}. Recently, the thermodynamic stability of these solutions has been investigated, and super-entropic black holes have been conjectured to be thermodynamically unstable \cite{Johnson2020,Cong2019}.
\\ In this framework, we observe a violation of the reverse isoperimetric inequality, implying that the black hole is super-entropic. Its thermodynamic stability has been explored computing the specific heats at constant pressure and volume. We discuss the results of these analysis in light of the conjecture on super-entropic black holes proposed in \cite{Johnson2020,Cong2019}.
\\The paper is structured as follows. In section~\ref{ch:EffectiveTheories} the model is presented and the effective second order theory is derived. In section~\ref{sec:TopologicalBHsolution} we report the hairy black hole solution together with its geometric characterization. Section~\ref{BH thermo A} is devoted to the black hole thermodynamics and the computation of the Euclidean action, while the violation of the reverse isoperimetric inequality and the thermodynamic stability analysis can be found in section~\ref{BH thermo B}. Finally, conclusions are drown in section~\ref{conslusions}.

\section{Effective theory}\label{ch:EffectiveTheories}
\noindent Let us consider the following generalization of the Holst action in vacuum\footnote{We work in geometric units in $G=c=1$.}
\begin{equation}
   I= \intgr \left[ f( \mathcal{R}+\mathcal{H}) - W(\gamma) \right],
   \label{action}
\end{equation}
where $\mathcal{R}=g^{\mu\nu}\mathcal{R}\indices{^{\rho}_{\mu\rho\nu}}$ is the Ricci scalar and the Riemann tensor is defined in terms of the connection $\Gamma\indices{^{\mu}_{\nu\rho}}$ (independent of the metric), as
\begin{equation}\label{Riemann tensor}
\tensor{\mathcal{R}}{^{\mu}_{\nu\rho\sigma}}=
\partial_{\rho}\tensor{\Gamma}{^{\mu}_{\nu\sigma}} - \partial_{\sigma}\tensor{\Gamma}{^{\mu}_{\nu\rho}} + \tensor{\Gamma}{^{\mu}_{\lambda\rho}}\tensor{\Gamma}{^{\lambda}_{\nu\sigma}} - \tensor{\Gamma}{^{\mu}_{\lambda\sigma}}\tensor{\Gamma}{^{\lambda}_{\nu\rho}}.
\end{equation}
The Holst term $\mathcal{H}$ is defined by
\begin{equation}
\mathcal{H}= - \frac{\gamma(x)}{2}\varepsilon^{\mu\nu\rho\sigma}\mathcal{R}_{\mu\nu\rho\sigma},
\end{equation}
and we promote the Immirzi parameter $\gamma$ to a scalar field with a potential term $W(\gamma)$. Here, we do not discuss in detail the effective mechanism able to generate such interaction term\footnote{This issue will be the object of a forthcoming work, along with the possibility of endowing the Immirzi parameter of dynamic from extended kinematical frameworks.}, so we just include the function $W(\gamma)$ in the action.

Now, by adopting a metric-affine formalism, we are implicitly assuming that the independent connection could be a priori characterized by non vanishing torsion and non-metricity tensors, defined as, respectively:
\begin{equation}
    T\indices{^{\lambda}_{\mu\nu}}\equiv\Gamma\indices{^{\lambda}_{\mu\nu}} - \Gamma\indices{^{\lambda}_{\nu\mu}}\qquad Q_{\mu\nu\rho}\equiv -D_{\mu}g_{\nu\rho},
\end{equation}
where $D_\mu$ stands for the general covariant derivative.
In the following, we will neglect non-metricity and just retain the anti-symmetric part of the connection. Even if this choice can seem a bit arbitrary, we are actually taking advantage of the invariance of the action \eqref{action} under the projective transformation
\begin{equation}
    \label{projective inv}
    \Gamma\indices{^{\rho}_{\mu\nu}} \rightarrow \Gamma\indices{^{\rho}_{\mu\nu}} + \delta^{\rho}_{\mu}\xi_{\nu},
\end{equation}
which can be exploited for simplifying the form of the connection, without affecting the dynamics \cite{Iosifidis:2018jwu,Iosifidis:2018zjj,Bejarano:2019zco, Iosifidis:2019fsh}. This is always attainable for the Lagrangian we are considering, where only the vector modes of the connection can be excited and 3-rank tensor states safely disregarded (see below for details in connection decomposition). In other words, \eqref{projective inv} constitutes a truly gauge symmetry for this kind of metric-affine models and in setting $Q_{\mu\nu\rho}=0$ we are just selecting a specific representation of \eqref{action}. 

We then introduce the contorsion tensor
\begin{equation}
K\indices{^{\mu}_{\nu\rho}} = \frac{1}{2}\left( T\indices{^{\mu}_{\nu\rho}} - T\indices{_{\nu}^{\mu}_{\rho}} -T\indices{_{\rho}^{\mu}_{\nu}} \right),
\end{equation}
which allows us to rewrite connection as
\begin{equation}\label{conn decomposition}
\Gamma\indices{^{\mu}_{\nu\rho}} = \bar{\Gamma}\indices{^{\mu}_{\nu\rho}} + K\indices{^{\mu}_{\nu\rho}},
\end{equation}
where $\bar{\Gamma}\indices{^{\mu}_{\nu\rho}} $ is the usual torsionless Levi-Civita connection for the metric $g_{\mu\nu}$.  Now, by standard methods (see e.g. \cite{Sotiriou2010}), it is possible to reformulate the theory in the Jordan frame. This can be done introducing an auxiliary field $\chi$ and rewriting the action as
\begin{equation}
    I= \intgr \left[ f(\chi) +f_\chi(\chi)( \mathcal{R}+\mathcal{H}-\chi) - W(\gamma) \right],
    \label{action aux}
\end{equation}
where a subscript denotes a derivative with respect to the argument. Provided\footnote{Actually, the condition for the second derivative to be non vanishing is not strictly necessary. It is sufficient to assume that $f$ be continuous and one-to-one, as shown in \cite{PhysRevD.75.023511}.} $f_{\chi\chi}\neq0$, variation with respect to $\chi$ yields the condition $\chi=\mathcal{R}+\mathcal{H}$, which reinserted into the action proves the equivalence with \eqref{action}. Then, introducing the scalaron field defined as $\phi\equiv f_\chi$, action \eqref{action aux} can be recast in the equivalent scalar-tensor theory
\begin{equation}\label{first order action}
I= \intgr \left[ \phi (\mathcal{R}+\mathcal{H}) -V(\phi) - W(\gamma) \right],
\end{equation}
where the potential is given by $V(\phi) = \phi \chi(\phi) - f(\chi(\phi))$ and $\chi(\phi)$ is obtained inverting the definition of the scalaron.

To find analytical solutions it is convenient to solve first the equations of motion for the independent connection. This can be accomplished writing the torsion tensor in terms of its independent components, i.e. the trace vector
\begin{equation}
T_{\mu}\equiv\tensor{T}{^{\nu}_{\mu\nu}},
\end{equation}
the pseudotrace axial vector
\begin{equation}
S_{\mu}\equiv \varepsilon_{\mu\nu\rho\sigma} T^{\nu\rho\sigma}
\end{equation}
and an anti-symmetric tensor $q_{\mu\nu\rho}=-q_{\mu\rho\nu}$, satisfying $\tensor{q}{^{\nu}_{\mu\nu}}=0$ and $\varepsilon^{\rho\nu\sigma\mu}q_{\nu\sigma\mu}=0$. In terms of these quantities the torsion tensor can be written as
\begin{equation}\label{TorsionSplittingFla}
T_{\mu\nu\rho} = \dfrac{1}{3}\left(T_{\nu}g_{\mu\rho}-T_{\rho}g_{\mu\nu}\right) +\dfrac{1}{6} \varepsilon_{\mu\nu\rho\sigma}S^{\sigma} + q_{\mu\nu\rho}.
\end{equation}
Substituting this into \eqref{conn decomposition}, one can write the action in terms of $T_{\mu}$, $S_{\mu}$ and $q_{\mu\nu\rho}$, as well as the metric tensor and the scalar fields. In particular, the Ricci scalar and the Holst term can be decomposed as, respectively \cite{Mercuri2006,Calcagni2009}:
\begin{align}
    &\mathcal{R}=\bar{R}+\frac{1}{24}S_\mu S^\mu-\frac{2}{3}T_\mu T^\mu-2\nabla_\mu T^\mu+\frac{1}{2}q_{\mu\nu\rho}q^{\mu\nu\rho}\\
    &\mathcal{H}=-\frac{\gamma(x)}{2}\leri{\nabla_\mu S^\mu+\frac{2}{3}T_\mu S^\mu+\frac{1}{2}\varepsilon^{\mu\nu\rho\sigma}q\indices{^\lambda_{\mu\nu}}q_{\lambda\rho\sigma}}
    \label{Holst decomposition}
\end{align}
where $\bar{R}$ and $\nabla_\mu$ are built from the Levi-Civita connection. Then, a straightforward computation of the equations of motion for the components of the torsion shows that there are solutions characterized by $q_{\mu\nu\rho}=0$ and
\begin{align}\label{eq: torsion components}
T_{\mu}&=\frac{3}{2\phi}\nabla_{\mu}\phi + \frac{3\gamma}{2(\gamma^2+1)}\nabla_{\mu}\gamma,\\
S_{\mu}&=-\frac{6}{(\gamma^2+1)}\nabla_{\mu}\gamma.
\end{align}
Therefore, the torsion acquires an effective dynamics sourced by the scalar field, and by plugging the above expressions back into the action, we finally obtain a scalar-tensor theory described by
\begin{align}\label{Holst JF tor solved}
  I&=\intgr \Bigl[ \phi \bar{R} +\frac{3}{2\phi} \nabla_{\mu}\phi\nabla^{\mu}\phi  - \frac{\phi}{2}\nabla_{\mu}\psi\nabla^{\mu}\psi \nonumber \\
  & -V(\phi) - W(\psi) \Bigr],
  \end{align}
where we defined the scalar field $\psi$ as
\begin{equation}\label{eq: Immirzi reparametrization}
\psi(x)\equiv\sqrt{3}\sinh^{-1}\gamma(x),
\end{equation}
and the potential $W$ has to be understood as a function of $\psi$ via the inversion of \eqref{eq: Immirzi reparametrization}.
We note that the transition to the Jordan frame results in the non-minimal coupling of the scalaron $\phi$, which turns out to multiply the Ricci scalar in the action. This is a peculiar feature of $f(\mathcal{R})$-like theories and it will have several implications in the thermodynamic treatment of Section~\ref{BH thermo A}.
We emphasize that when the Immirzi field relaxes to a constant $\gamma_0$, the model \eqref{Holst JF tor solved} boils down to the standard Palatini $f(\mathcal{R})$ gravity in the presence of the additional cosmological term, due to the potential terms (see discussion in section~\ref{sec:TopologicalBHsolution}). Furthermore, we stress the fact that choosing from the very beginning models of the type $f(\mathcal{R})+\mathcal{H}$, with the Holst term outside the argument of the function $f(\cdot)$, does not really alter the form of \eqref{Holst JF tor solved}. In this case, indeed, it suffices to redefine the scalar field $\psi$ as $\psi=\sqrt{3}\sinh^{-1}(\gamma/\phi)$ to find again \eqref{Holst JF tor solved}, the only difference consisting in the function $W$, which now also depends on the field $\phi$. 
Now, before dealing in detail with the equations of motion stemming from \eqref{Holst JF tor solved}, we observe that the theory is safely devoid of ghost instabilities, even if the kinetic term of the scalaron appears with the wrong sign. In the event of non-minimal coupling, indeed, such a sign is not sufficient to determine the presence of ghost modes and, as discussed in \cite{Fujii2003}, it has to be evaluated in the so called Einstein frame, defined by the metric rescaling $\tilde{g}_{\mu\nu}=\phi g_{\mu\nu}$. In this case \eqref{Holst JF tor solved} is recast as
\begin{align}\label{Holst JF tor solved E}
  I_E&=\intgrE \left[ \tilde{R}  - \frac{1}{2}\tilde{g}^{\mu\nu}\nabla_{\mu}\psi\nabla_{\nu}\psi- U(\phi,\psi)\right],
  \end{align}
with $U(\phi,\psi)\equiv (V(\phi)+W(\psi))/\phi^2$, and we see that the only dynamical scalar field is the reparametrized Immirzi field, whose kinetic term has the correct sign. The scalaron $\phi$ is not dynamical and its evolution is entirely determined  by the field $\psi$, as it is evident by varying \eqref{Holst JF tor solved E} with respect to it, i.e.
\begin{equation}\label{eq: struct eq Immirzi}
2V(\phi)-\phi \frac{dV}{d\phi}= -2W(\psi),
\end{equation}
which is nothing but a generalization of the so-called structural equation of the standard Palatini $f(\mathcal{R})$ theory in vacuum, where it reads as \cite{Olmo2011}
\begin{equation}\label{eq:struct eq Pal f(R)}
2V(\phi)-\phi \frac{dV}{d\phi} = 0.
\end{equation}
Now, coming back to the action \eqref{Holst JF tor solved}, we report the missing field equations for the metric and the scalar $\psi$, which read, respectively
\begin{align}
G_{\mu\nu}&=\frac{1}{\phi}\left( \nabla_{\mu}\nabla_{\nu}\phi - g_{\mu\nu}\Box\phi \right)-\frac{3}{2\phi^2}K_{\mu\nu}(\phi)+\frac{1}{2}K_{\mu\nu}(\psi)\nonumber\\
&-\frac{1}{2\phi}\leri{V(\phi)+W(\psi)}, \label{equation metric}\\
\Box\psi&=-\nabla^{\mu}\psi\nabla_{\mu}\text{ln}\phi
+\frac{1}{\phi}\frac{dW}{d\psi},
\label{equation psi}
\end{align}
where $\Box=\nabla_{\mu}\nabla^{\mu}$ is the d'Alambert operator built from the Levi-Civita connection and
\begin{equation}
K_{\mu\nu}(\cdot)\equiv\nabla_{\mu}(\cdot)\nabla_{\nu}(\cdot)-\frac{1}{2}g_{\mu\nu}\nabla^{\rho}(\cdot)\nabla_{\rho}(\cdot).
\end{equation}
We point out that \eqref{eq: struct eq Immirzi} can be still obtained from \eqref{Holst JF tor solved}, with a bit of additional effort, and we do not discuss it. We just stress that, in contrast with \eqref{eq:struct eq Pal f(R)}, in our case \eqref{eq: struct eq Immirzi} establishes an algebraic relation between $\phi$ and $\psi$, and the scalaron can acquire in vacuum a non-trivial profile, as opposed to \eqref{eq:struct eq Pal f(R)} where it is compelled to relax to a constant.

\section{Topological hairy black hole}\label{sec:TopologicalBHsolution}
No-hair theorems prevent the existence of black hole solutions with hair, namely scalar fields with a non-trivial functional form. These can be either of the primary or secondary kind, depending on the presence or absence, respectively, of a related independent charge (see \cite{Sotiriou_2015} for details). However, there are several ways to evade no-hair theorems by violating some of the hypothesis on which they stand, allowing the possibility of hairy black holes in scalar-tensor theories. Among these, one may relax the asymptotic flatness assumption. In this paper,  we follow this possibility to work out an analytical solution describing a hairy, asymptotically Anti-de Sitter, topological black hole. The hair are of the secondary kind and the solution generalizes the results of \cite{Martinez2004} to the case of non-minimal coupling and reduces to it for appropriate values of the parameters characterizing the model. This is chosen to be a generalization of the Starobinsky model \cite{STAROBINSKY198099} with the inclusion of a cosmological constant term\footnote{The prefactor is chosen for later convenience}:
\begin{equation}\label{eq:f(R)Model}
f(\chi) = \frac{1}{1+8\alpha \Lambda} \left( \chi + \alpha \chi^2 - 2\Lambda \right),
\end{equation}
for a general argument $\chi$. The metric of the solution is given by
\begin{equation}\label{solution metric}
ds^2 = \Omega(r) \left[ -h(r) dt^2 + h^{-1}(r)dr^2 + r^2 d\sigma^2 \right],
\end{equation}
where
\begin{equation}
h(r)= -\left(1+\frac{m}{r}\right)^2 +\frac{r^2}{l^2},
\end{equation}
with $\Lambda=-3/l^2<0$, and $d\sigma$ is the line element of a 2-surface of constant negative curvature $\Sigma$. It has hyperbolic topology\footnote{A solution with trivial spherical topology exists as well, but it has no physical interpretation since the lapse function has no roots, making the origin a naked singularity.} and genus $g\geqslant 2$, with area $\sigma=4\pi(g-1)$ \cite{Vanzo1997,Birmingham1999,Smith1997}. The Immirzi field surrounds it with a secondary scalar hair, expressed by
\begin{equation}\label{solution Immirzi field}
\psi(r) = \psi_0  + \sqrt{12} \, \text{arctanh}\left( \frac{m}{r+m} \right),
\end{equation}
where $\psi_0$ is a constant. The conformal factor reads
\begin{equation}
\Omega(r)=\frac{r(r+2m) + 48 \alpha m^2/l^2}{(r+m)^2},
\end{equation}
where the parameter $\alpha$ characterizes the Jordan frame potential corresponding to model \eqref{eq:f(R)Model}, given by
\begin{equation}\label{eq:JFPotential}
V(\phi) = \frac{\left( \phi-1 \right)^2}{4\alpha} + 2\Lambda \phi^2.
\end{equation}
Note that $\Lambda$ does not enter the theory as a true cosmological constant, namely a constant term added to the Jordan frame action. Indeed, we see that the actual constant term in \eqref{Holst JF tor solved} comes from \eqref{eq:JFPotential} and reads $1/(4\alpha)$. However, it is $\Lambda$ that rules the asymptotic behaviour of the metric and, primarily, of the Ricci scalar of the metric \eqref{solution metric}, given by
\begin{equation}
\bar{R} \sim -\frac{12}{l} + O\left( \frac{1}{r^2} \right),
\end{equation}
which tells us that the spacetime is asymptotically Anti-de Sitter space with radius $l$.\\
With the choice \eqref{eq:f(R)Model} the Immirzi field potential is given by
\begin{equation}
W(\psi) = \frac{4\Lambda}{ \text{csch}^2\left( \frac{\psi-\psi_0}{\sqrt{12}} \right)-16\alpha\Lambda}.
\end{equation}
It is characterized by the negative mass term $\frac{d^2W}{d\psi^2}{\big |}_{\psi_0}=-2/l^2$, which satisfies the Breitenlohner-Freedman bound for the  stability in Anti-de Sitter space \cite{Breitenlohner1982,Mezincescu1985}.
 The scalar field $\phi$ is determined by the structural equation \eqref{eq: struct eq Immirzi}, which  yields
\begin{equation}\label{eq:PhiPsiRelation}
\phi = 1+4\alpha W(\psi).
\end{equation}
For $\alpha=0$ and $\psi_0=0$ the above solution reduces to the solution found in \cite{Martinez2004}. Even if the potential $V(\phi)$ is singular in $\alpha=0$, the limit can be safely taken a priori in \eqref{eq:f(R)Model}, 
which reduces to $f(\chi)=\chi-2\Lambda$, yielding $V=2\Lambda$.

Now, let us study the horizon structure of the black hole and the behaviour of the scalar fields.
In addition to the origin $r=0$, there are two other curvature singularities $r^{\begin{tiny}
\Omega
\end{tiny} }_{\pm}=-m\pm\sqrt{m^2\left( 1-48\alpha/l^2 \right)}$, corresponding to the roots of the conformal factor $\Omega(r)$, in which the scalars of curvature diverge.
The coordinate singularities, instead, are located where the metric function $h(r)$ vanishes. One of its roots is always negative, while the others are
\begin{align}
    r_e&=\frac{l}{2}\left( 1+\sqrt{1+\frac{4m}{l}} \right),\\
    r_+&=\frac{l}{2}\left( 1-\sqrt{1+\frac{4m}{l}} \right),\\
    r_-&=\frac{l}{2}\left( -1+\sqrt{1-\frac{4m}{l}} \right).
\end{align}
For $m>0$, the only positive real root is $r_e$. For negative mass parameter we distinguish two cases. For $-l/4<m<0$ there are three positive real roots, namely $0<r_-<r_+<r_e$, and for $m<-l/4$, the only positive real root is $r_-$.\\
The value of the parameters $m$ and $\alpha$ determine if the solution has a black hole interpretation or if it consists of a naked singularity. In particular, it can be shown that:
\begin{itemize}
    \item For $\alpha \geqslant l^2/48$ one has that $r^{\Omega}_{\pm}$ become complex and the only curvature singularity is at the origin and it is always hidden behind event horizons at $r_e$ or $r_-$. In this case the spacetime is a black hole for every value of $m$.
    \item For $0\leqslant\alpha<l^2/48$, the curvature singularities are hidden only for $m>-l/4$.
    \item For $\alpha<0$ the mass parameter must satisfy \\ $m_-<m<m_+$, where
\begin{align}
    m_-&=-\frac{l^2 \sqrt{l^2-48 \alpha }}{2 l \left(\sqrt{l^2-48 \alpha }+l\right)-48 \alpha }\geqslant -\frac{l}{4},\\
    m_+&= \frac{l^2 \sqrt{l^2-48 \alpha }}{\left(l-\sqrt{l^2-48 \alpha }\right)^2}.
\end{align}
\end{itemize}
As we will see in the next section, $m$ is related to the mass-energy of the black hole. The existence of the upper bound $m_+$ implies that an increasing of the black hole mass would result in developing a naked singularity. To exclude this possibility we restrict in the following the parameter $\alpha$ to positive values, ruling out models described by $\alpha<0$.\\
In the limiting case $m=m_c\equiv -l/4$, the metric function $h(r)$ has two positive roots, the greater one being $r=r_c\equiv l/2$.
This critical configuration will be important in the computation of the Euclidean action to which the next section is dedicated.
In the following, thermodynamic reasons will constrain the mass parameter to obey $m>m_c$, therefore, in all cases, the outer event horizon is located at $r_e$.\\
We now come to the scalar fields. The scalar field $\phi$ has two poles coincident with the roots of $\Omega(r)$. If $m$ is taken to be in the above mentioned range, they are located at negative or complex radius or hidden by the event horizon, depending on the value of $\alpha$. The field is regular on and outside the event horizon, with a radial profile monotonically increasing (decreasing) from $\phi(r_e)$ to $1$, which is reached asymptotically as $r\rightarrow \infty$, for $\alpha>0$ $(\alpha<0)$.
Taking into account reparametrization \eqref{eq: Immirzi reparametrization}, the Immirzi field profile is given by
\begin{equation}
\gamma(r) = \frac{e^{\psi_0/\sqrt{3}}(r+2m)^2-e^{-\psi_0/\sqrt{3}}r^2}{2r(r+2m)}.
\end{equation}
It depends only on the mass parameter and on its asymptotic value $\gamma_0 \equiv \text{sinh}(\frac{\psi_0}{\sqrt{3}})$, which is reached as $r\rightarrow\infty$. Thus, in the asymptotic region the Immirzi field relaxes to a constant value and $\phi\rightarrow 1$, together with $W\rightarrow 0$ and $V\rightarrow 2\Lambda$, implying that, asymptotically, the theory reduces to GR with a constant Immirzi parameter, namely to the usual formulation of LQG with a cosmological constant $\Lambda$. In this limit the bare cosmological constant present in \eqref{Holst JF tor solved} cancels with the $-1/(4\alpha)$ term coming from $W(\psi)$.

\section{Black Hole Thermodynamics}
\subsection{Computation of the Euclidean action}\label{BH thermo A}
We study the thermodynamic properties of the black hole solution of section~\ref{sec:TopologicalBHsolution} via the Euclidean path integral methods. The usual procedure \cite{PhysRevD.15.2752,doi:10.1142/S0218271814300237} consists in starting from the gravitational partition function and defining the thermodynamic partition function $Z(\beta)$, via a Wick rotation to imaginary time $t\rightarrow i \tau$ and imposing periodic boundary conditions on the Euclidean time. The period $\beta$ can be identified with the inverse temperature, and a saddle point approximation around a classical solution allows to write $Z(\beta) \approx e^{-I(\beta)}$, where $I(\beta)$ is the on-shell action in Euclidean signature. Then, usual thermodynamic relations hold, as for instance
\begin{equation}\label{eq: thermodynamics relation}
I=S-\beta M,
\end{equation}
which relates the Euclidean on-shell action with the mass-energy $M$ and entropy $S$ of the black hole.

After the Wick rotation to imaginary time, the Euclidean metric reads
\begin{equation}
ds_E^2 = \Omega(r) \left[ h(r)d\tau^2 +h^{-1}(r) dr^2 + r^2 d\sigma^2 \right].
\end{equation}
As usual, the regularity of the metric at the horizon must be required by fixing the Euclidean time periodicity. For $r\approx r_e$, the near horizon metric is
\begin{equation}
ds^2_E = d\tilde{r}^2 + \frac{h'(r_e)^2}{4} \tilde{r}^2 d \tau^2 + \Omega(r_e)r_e^2d\sigma^2,
\end{equation}
where a prime denotes derivatives with respect to $r$ and we have defined a new radial coordinate as
\begin{equation}
\tilde{r} = 2\sqrt{\frac{\Omega(r_e)(r-r_e)}{h'(r_e)}}.
\end{equation}
The $\tilde{r}-\tau$ section of the metric is just flat space in polar coordinates provided that the conical singularity at the origin is removed by identifying the Euclidean time with an angular coordinate of period $\beta$ given by
\begin{equation}
\beta=\frac{4\pi}{h'(r_e)}=\frac{2\pi l^2}{2 r_e-l}.
\end{equation}
The black hole temperature is identified with the inverse of the period, namely
\begin{equation}\label{temperature}
T=\frac{1}{2\pi l}\left( \frac{2 r_e}{l}-1 \right).
\end{equation}
We see that, to ensure the positivity of the temperature, the horizon radius must satisfy $r_e>r_c\equiv l/2$, or, in terms of the mass parameter, $m>m_c\equiv-l/4$. The solution identified by $m_c$ and $r_c$ corresponds to the limiting case mentioned in the previous section. An analogue configuration was described in \cite{Vanzo1997,Birmingham1999}, where such a critical configuration corresponds to the minimum value of the mass parameter, which still allows a black hole interpretation of the solution. For smaller masses, a naked singularity develops.
In the present case, the same holds only if no restrictions on the model are imposed. Indeed, the curvature singularities can always be concealed restricting $\alpha$ to be greater than $l^2/48$, allowing for a black hole interpretation for every value of $m$.

When studying black hole thermodynamics via the computation of the Euclidean action one has to pay particular attention to two problems: the first is that the action does not generally yield a well-posed variational principle and the second is that its on-shell value is usually infinite. The first issue can be solved with the inclusion of a GHY-like surface term \cite{Dyer2009}, proportional to the extrinsic curvature of the boundary and given by
\begin{equation}\label{GHY term}
    I_{GHY}= \frac{1}{8\pi}\int_{\partial\mathcal{M}}d^3x\sqrt{|{}^{(3)}g|}\phi K,
\end{equation}
where ${}^{(3)}g$ is the determinant of the induced metric on the boundary $\partial\mathcal{M}$ and $K$ the trace of its extrinsic curvature.

The non-minimal coupling between $\phi$ and the Ricci scalar in \eqref{Holst JF tor solved} is responsible for the discrepancy between \eqref{GHY term} and the usual GHY term, in which $\phi$ is absent. In this way, the variation of \eqref{GHY term} exactly cancels non vanishing boundary contributions arising from varying the first term of \eqref{Holst JF tor solved}.

Note that the first order action \eqref{first order action} yields a well-posed variational principle without the need of additional boundary terms. However, the correct equivalent second order action, namely the one yielding an equivalent set of field equations via a well-posed variational principle, is not simply \eqref{Holst JF tor solved}, whose variation would give rise to unwanted non-vanishing boundary terms arising from the $\phi\bar{R}$ term in the action, but should be instead completed with the inclusion of \eqref{GHY term}.

We address the second issue via the counter-terms method \cite{Myers1999,Gegenberg2003} which consists in adding counter-terms to the action which are surface integrals depending on the induced metric on the boundary and, possibly, on the scalar fields of the theory. The method can also be applied without specifying the explicit expression of the counter-terms \cite{Martinez2004}.
In this section we follow the latter approach, generalizing the treatment of \cite{Martinez2004} to the case of non-minimal coupling. The explicit covariant expression of the counter-terms will be nevertheless shown at end of this section.

Let us first rewrite the Euclidean metric as
\begin{equation}
ds^2_E = N^2(r)f^2(r) d\tau^2 + f^{-2}(r) dr^2 + \rho^2(r)d\sigma^2,
\end{equation}
where the new metric functions are related to the previous ones by
\begin{equation}
N=\Omega, \qquad\qquad f^2=\frac{h}{\Omega}, \qquad\qquad \rho^2=\Omega\, r^2.
\end{equation}
The Euclidean version of action \eqref{Holst JF tor solved} can be written in Hamiltonian formalism as
\begin{equation}\label{eq: Euclidean action}
I = -\frac{\beta\sigma}{8\pi}\int_{r_e}^{\infty}dr N H + B,
\end{equation}
having integrated over $\tau$ and the base manifold $\Sigma$. Here, $B$ represents an appropriate boundary term, whose role is twofold: on one hand it makes the variational principle well-posed and, on the other hand, it cures the divergence of the on-shell action. The Hamiltonian reads
\begin{align}
H=&\rho^2 \left\lbrace \phi\left[ \frac{{f^2}'\rho'}{\rho}+ \frac{2f^2\rho''}{\rho}+\frac{(1+f^2\rho'^2)}{\rho^2} \right] \right. \nonumber\\ &\left.-\frac{3}{4\phi}f^2\phi'^2+\frac{\phi}{4}f^2\psi'^2+ \frac{V(\phi)+W(\psi)}{2}\right.\nonumber+\\
&\left.+\frac{{\rho^2}'f^2\phi'}{\rho^2} +\frac{ {f^2}'\phi'}{2}+f^2 \phi'' \right\rbrace.
\end{align}
The third line shows additional contributions arising from the non-minimal coupling which are absent in \cite{Martinez2004}.

In the expressions above the terms involving the momenta and the shift vector are absent since the solution is static and spherically symmetric. Moreover, one should also include in the action an additional term proportional to the structural equation which is known to manifest itself as a secondary constraint in Hamiltonian formalism \cite{Olmo2011}. However, this term would not contain derivatives of the fields with respect to $r$ and thus it is irrelevant in the following calculations.

Now, the Hamiltonian vanishes on shell, thus the only contribution to the Euclidean action comes from the boundary term. The latter can be computed varying the action with respect to the metric functions and the scalar fields as
\begin{equation}
\delta I = -\frac{\beta\sigma}{8\pi}\int_{r_e}^{\infty}dr N \delta H + \delta B.
\end{equation}
Then, one can choose $\delta B$ to be such that it cancels the boundary terms arising from the variation of the Hamiltonian, that is
\begin{equation}\label{eq:BoundaryTermVariation}
\delta B = \delta B_g + \delta B_\phi + \delta B_\psi,
\end{equation}
where
\begin{align}\label{deltaBg}
\delta B_g & = \frac{\beta\sigma}{8\pi}\left[ \left(N\rho\phi \rho' + \frac{1}{2}N\rho^2{\phi}' \right)\delta f^2\right.\nonumber\\
&\left.-\left( 2N'\rho\phi f^2+N\rho\phi {f^2}' \right)\delta \rho  +2N\rho\phi f^2 \delta \rho' \right]_{r_e}^{\infty},\\
\label{deltaBphi}\delta B_\phi &=\frac{\beta\sigma}{8\pi}\left[ -\left( \frac{3}{2\phi}N\rho^2f^2\phi' +N'\rho^2f^2+\frac{1}{2}N\rho^2{f^2}'\right)\delta\phi \right.\nonumber\\
&\left.+ N\rho^2f^2\delta \phi' \right]_{r_e}^{\infty},\\
\delta B_{\psi} &= \frac{\beta\sigma}{8\pi}\left[ \frac{1}{2}\phi N \rho^2 f^2 \psi' \delta\psi \right]_{r_e}^{\infty}.
\end{align}
The variation of the fields at infinity are
\begin{align}
\delta f^2{\big |}_{\infty} &=\left[\frac{2 m \left(l^2-48 \alpha \right)}{l^4}+\frac{6 m^2 \left(48\alpha-l^2  \right)-2l^4}{l^4r}\right.\nonumber\\
&\left.+O\left(\frac{1}{r^2}\right)\right]\delta m,\\
\delta \rho{\big |}_{\infty} &=\left[ \frac{m \left(48 \alpha-l^2\right)}{l^2r}+\frac{3 m^2 \left(l^2-48 \alpha \right)}{l^2 r^2}\right.\nonumber\\
&\left.+O\left(\frac{1}{r^3}\right) \right]\delta m,\\
\delta \phi{\big |}_{\infty} &=\left[ -\frac{96 \alpha  m}{l^2 r^2}+\frac{288 \alpha  m^2}{l^2 r^3}+O\left(\frac{1}{r^4}\right) \right]\delta m,\\
\delta \psi{\big |}_{\infty} &=\left[ \frac{2 \sqrt{3}}{r}-\frac{4 \sqrt{3} m}{r^2}+O\left(\frac{1}{r^3}\right) \right]\delta m.
\end{align}
Substituting in \eqref{eq:BoundaryTermVariation} yields
\begin{equation}
\delta B {\big |}_{\infty} = -\frac{2\beta\sigma}{8\pi}\delta m + O\left( \frac{1}{r^2} \right).
\end{equation}
Therefore, the boundary term at infinity can be read off to be
\begin{equation}\label{B at infinity}
B {\big |}_{\infty} = -\frac{2\beta\sigma}{8\pi} m.
\end{equation}
To compute the boundary term at the horizon, let us first notice that $f^2(r_e)=h(r_e)/\Omega(r_e)=0$, which implies $\delta B_{\psi} {\big |}_{r_e}=0$ and simpler expressions for \eqref{deltaBg} and \eqref{deltaBphi} when computed at $r_e$. Then, one can use the following relations
\begin{align}
\delta \rho{\big |}_{r_e} &= \delta \rho(r_e)-\rho'{\big |}_{r_e}\delta r_e,\\
\delta f^2{\big |}_{r_e} &= -{f^2}'{\big |}_{r_e}\delta r_e,\\
\delta \phi{\big |}_{r_e} &= \delta \phi(r_e)-\phi'{\big |}_{r_e}\delta r_e,
\end{align}
to compute the variation of the boundary term at the horizon as
\begin{align}
\delta B{\big |}_{r_e} &= -\frac{\beta\sigma}{16\pi}\left[ N\phi {f^2}'\delta \rho^2(r_e) + N{f^2}' \rho^2\delta\phi(r_e) \right]\nonumber\\
& =-\frac{\beta\sigma}{16\pi} \, N{f^2}'{\big |}_{r_e} \delta\left( \phi(r_e)\rho^2(r_e) \right).
\end{align}
Recalling the definition of the Euclidean time period,
\begin{equation}
N{f^2}'{\big |}_{r_e} = h' {\big |}_{r_e} = \frac{4\pi}{\beta},
\end{equation}
the result can be written as
\begin{equation}
\delta B{\big |}_{r_e} = -\frac{\sigma}{4}\delta\left( \phi(r_e) \rho^2(r_e) \right),
\end{equation}
which leads to the boundary term at the horizon
\begin{equation}
B{\big |}_{r_e}=-\frac{\sigma}{4}\phi(r_e) \rho^2(r_e).
\end{equation}
The addition of the two contributions gives
\begin{equation}
I = -\frac{\beta \sigma}{4\pi} m + \phi(r_e) \frac{\sigma \rho^2(r_e)}{4}.
\end{equation}
By comparing this with \eqref{eq: thermodynamics relation} one finds that the mass-energy and entropy of the black hole are given, respectively, by
\begin{equation}
M=\frac{\sigma m}{4\pi},
\end{equation}
\begin{equation}
S=\phi(r_e) \frac{A}{4},
\end{equation}
where 
\begin{equation}\label{horizon area}
    A=\sigma \rho^2(r_e)=\sigma \Omega(r_e)r_e^2
\end{equation}
is the horizon area.

Now, recalling that the only dynamical scalar field is $\psi$, and that this is algebraically related to $\phi$ by \eqref{eq:PhiPsiRelation}, one concludes that the Immirzi field modifies the expression of the entropy, from the standard expression $A/4$ to
\begin{equation}\label{entropy}
S=\left[1+4\alpha W(\psi_e)\right]\frac{ A}{4},
\end{equation}
where $\psi_e$ is the Immirzi field computed at the black hole event horizon. Note that for $m=0$ the black hole has zero mass but non-vanishing entropy $S(m=0)=\sigma l^2/4$. This is consistent with results regarding AdS topological black holes with no hair \cite{Vanzo1997,Birmingham1999}.

Such a modification of entropy with respect to the standard area law is expected since the presence of non-minimal coupling and it is consistent with other derivations in similar contexts \cite{Faraoni2010}.

In the calculation performed in this section the explicit definition of the boundary term $B$ is never specified. However, its covariant expression can also be derived \cite{Myers1999,Gegenberg2003}. In our case we obtain that the above results are reproduced starting from the following finite, well-posed, action
\begin{equation}
    I+I_{GHY}+I_{ct}^1+I_{ct}^2+I_{ct}^{\psi},
\end{equation}
where we added the following counter-terms to \eqref{Holst JF tor solved}:
\begin{align}\label{counter terms 1}
    I_{ct}^1 &=-\frac{1}{8\pi}\int_{\partial \mathcal{M}}d^3x\sqrt{|{}^{(3)}g|}\frac{2}{l}\phi\sqrt{\phi},\\\label{counter terms 2}
    I_{ct}^2&=-\frac{1}{8\pi}\int_{\partial \mathcal{M}}d^3x\sqrt{|{}^{(3)}g|}\frac{l}{2}\sqrt{\phi} \;\indices{^3}R,\\\label{counter terms 3}
    I_{ct}^{\psi}&=\frac{1}{16\pi}\int_{\partial \mathcal{M}}d^3x\sqrt{|{}^{(3)}g|}\frac{\phi\sqrt{\phi}}{6l}\left[ \frac{2l(\psi-\psi_0)}{\sqrt{\phi}}n^{\mu}\partial_{\mu}\psi\right.\nonumber\\
    &\left.-\left(\psi-\psi_0\right)^2
    \right],
\end{align}
as well as the surface term \eqref{GHY term}. In the expressions above $\indices{^3}R$ is the three dimensional Ricci curvature of the boundary metric and $n^{\mu}$ the unit normal to the boundary. In the limit $\phi\rightarrow 1$, these counter-terms reduce to the ones reported in \cite{Gegenberg2003} for a minimally coupled scalar field.

Contrary to what happens in the absence of additional scalar fields or for localized distributions of matter \cite{Myers1999} with radial fall off $\sim r^{-3/2+\varepsilon}$ at infinity, the counter-terms \eqref{counter terms 1}, \eqref{counter terms 2} and \eqref{counter terms 3} explicitly depend on the scalar fields. The reason is the slower fall off of $\psi\sim\psi_0+2\sqrt{3}m r^{-1} + O(r^{-2})$ with respect to localized distributions of matter. The resulting back-reaction on the metric requires the counter-terms to depend also on the scalar fields in order to properly cancel divergences.
A different asymptotic behaviour of the scalar fields would lead to different counter-terms, as pointed out in \cite{Gegenberg2003} (see also \cite{PhysRevD.62.124002,Berg_2002,DeHaro2001}, where the same issue is analysed in three and higher dimensions).

The boundary term $B$ contains contributions coming both from the surface term \eqref{GHY term} and from the above counter-terms. Moreover, being inserted in the Hamiltonian version of the action, it actually contains also boundary terms arising from the Gauss-Codazzi relation, used in the spacetime splitting procedure. For this reason one cannot directly compare \eqref{B at infinity} with the asymptotic expansion of the above counter-terms, which are inserted at the Lagrangian level.

Finally, we note that expression \eqref{entropy} is consistent with the one obtained applying Wald's formula \cite{PhysRevD.48.R3427} for the entropy to the original first order action \eqref{first order action}, notwithstanding the presence of torsion in the theory (see \cite{CHAKRABORTY2018432} for a discussion on Wald's entropy in models with torsion).

\subsection{Reverse Isoperimetric Inequality violation}\label{BH thermo B}

Black hole thermodynamics has been studied in the context of the extended phase space approach \cite{Kubiznak2017}, where the cosmological constant is interpreted as a thermodynamic pressure given by
\begin{equation}
    P=-\frac{\Lambda}{8\pi },
\end{equation}
which is positive for asymptotically AdS spacetimes.

The corresponding conjugate quantity, the thermodynamic volume $V$, is computed via the first law of thermodynamics, which reads
\begin{equation}
    dM = TdS + VdP,
\end{equation}
where $M$ is interpreted as enthalpy rather than internal energy. In the present case, substituting the expressions for the other thermodynamic variables into the above relation yields
\begin{equation}\label{volume}
    V =\frac{\sigma l^2}{3}(l+3m)=\frac{\sigma}{3}\left(3lr_e^2 -3l^2r_e+l^3\right).
\end{equation}
This thermodynamic volume does not coincide with the geometric volume defined as $V=\sigma (r_e\sqrt{\Omega(r_e)})^3/3$, as it happens for solutions more complex than the Schwarzschild-AdS case \cite{PhysRevD.84.024037,PhysRevLett.115.031101,PhysRevLett.117.131303}. However, it has some common properties: it is a positive definite increasing monotonic function of $r_e$ (for $r>l/2$) and it is proportional to the genus $g$ of the horizon. It attains its minimum value $V_{min}=\sigma l^3/12$ at $r_e=r_c$. One can also verify that the Smarr formula holds, i.e.
\begin{equation}\label{smarr formula}
    M = 2(TS - PV).
\end{equation}
The physical meaning and properties of the thermodynamic volume have been studied extensively in literature and in \cite{Dolan2013} it was conjectured that for every asymptotically AdS black hole, the reverse isoperimetric inequality (RII) holds, namely that $\mathcal{I}\geqslant 1$, where
\begin{equation}
    \mathcal{I} = \left(\frac{(d-1) V}{\omega_{d-2}^{(k)}} \right)^{\frac{1}{d-1}}\left(\frac{\omega_{d-2}^{(k)}}{A}\right)^{\frac{1}{d-2}},
\end{equation}
for arbitrary dimension $d$ and generalized unit volume $\omega_{d-2}^{(k)}$ of the $d-2$ dimensional base manifold of constant curvature $k$. The conjecture was originally motivated by the observation that all known solutions seemed to satisfy the inequality. However, an increasing number of counterexamples have been found for which the conjecture is violated.
\\ The lower bound $\mathcal{I}=1$ is saturated by the Schwarzschild-Anti de Sitter (SAdS) black hole implying that, according to the conjecture, this would be the solution maximising the entropy for a given thermodynamic volume. For this reason solutions violating the conjecture have been called super-entropic black holes. For a given $V$ they allow for a greater area and therefore\footnote{This implication is trivially true when the entropy is given by $S=A/4$. However, for more complex cases, as the one considered here, it should be verified explicitly as we do in the following.} a greater entropy than the SAdS case. Examples include black holes with non-compact horizons, Lifshitz black holes, three-dimensional black holes \cite{PhysRevD.95.046002,PhysRevD.92.044015,PhysRevD.89.084007,PhysRevLett.115.031101}. Violations for hairy black holes with planar ($k=0$) horizons in four dimensions were also observed in \cite{Feng2017}.

The black hole solution of section~\ref{sec:TopologicalBHsolution} represents thus a new kind of super-entropic black hole. Indeed, in the present case, i.e. for $d=4$, $k=-1$ and $\omega_2^{(-1)}=\sigma$, one has
\begin{equation}
   \mathcal{I}= \frac{\left(3lr_e^2 -3l^2r_e+l^3\right)^{\frac{1}{3}}}{\sqrt{l(2r_e-l)+\frac{48\alpha}{l^2}(r_e-l)^2}}.
\end{equation}
We observe a violation of the RII in almost all parameter space. In particular, for $\alpha\geqslant l^2/(24\sqrt[3]{2})$ the violation occurs for every $r_e>r_c$, namely for every $T>0$ and $V>V_{min}$. Therefore, in this case, the black hole is always super-entropic.
\\ For $\alpha=0$, namely for the MTZ black hole \cite{Martinez2004}, the conjecture is satisfied and $\mathcal{I}\geqslant 1$, the inequality being saturated for $r_e=l$, which corresponds to pure AdS. For the sake of clarity we postpone the analysis of the case $0<\alpha<l^2/(24\sqrt[3]{2})$ to the end of this section, but we anticipate that all the conclusions reached in the following remain valid. 
\\ Now, whenever entropy and area are simply proportional, as it occurs for $S=A/4$, it is trivially true that the inequality $\mathcal{I}<1$ implies that the entropy can be larger than the bound saturated by SAdS. In our case instead, the relation between $S$ and $A$ is given by \eqref{entropy}. Does a violation of the RII still imply that the black hole is super-entropic? When $\mathcal{I}<1$, and at fixed volume, the black hole can have a larger area than the one of a SAdS black hole.
Moreover, solving \eqref{horizon area} for $r_e(A)$ (choosing the positive branch, for which $r_e>0$ for $A>0$) and then substituting it in the definition of the entropy, yields
\begin{equation}
    S(A) = \frac{\sigma l^2}{96\alpha}\left(24\alpha-l^2+\sqrt{l^2(l^2-48\alpha)+\frac{48\alpha}{\sigma}A}\right),
\end{equation}
which is a monotonically increasing function of $A$. Therefore, a violation of the RII still implies that the black hole is super-entropic.

In \cite{Johnson2020} this super-entropic behaviour was shown to be related to a thermodynamic instability expressed by a negative specific heat at constant volume $c_V$. Then, in \cite{Cong2019} exotic BTZ black holes were analysed showing that the RII can be violated even when $c_V>0$. However, the authors proved also that, whenever $c_V>0$, the specific heat at constant pressure $c_P$ becomes negative, still signalling a thermodynamic instability.

\begin{figure}[tbp]
\centering
\includegraphics[width=0.51\textwidth]{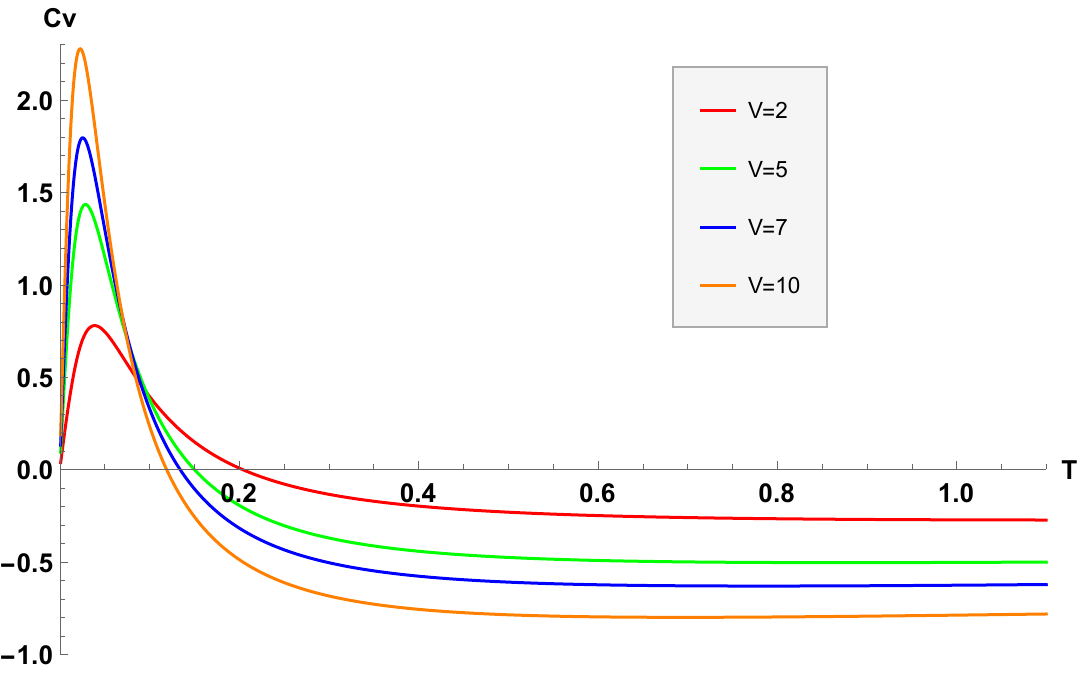}
\caption{\label{fig:cV(T)} Specific heat at constant volume $c_V$ as a function of $T$ for different values of $V$.}
\end{figure}

The black hole presented in section~\ref{sec:TopologicalBHsolution} is halfway between the two since $c_P$ is always positive, as in \cite{Johnson2020}, but there are black hole configurations violating the RII for which $c_V>0$. In spite of that, we concluded that super-entropic black holes are always thermodynamically unstable. To see this, we first note that, comparing \eqref{temperature} and \eqref{entropy}, the following relation between entropy and temperature can be derived:
\begin{equation}\label{relaz S(P,T)}
    S = \frac{\sigma\pi}{2}\left(\frac{3}{8\pi P}\right)^{\frac{3}{2}}T.
\end{equation}
Then, $c_P$ can be computed as
\begin{equation}\label{cp}
    c_P= T \frac{\partial S}{\partial T}{\Bigg |}_{P} = \frac{\sigma\pi}{2}\left(\frac{3}{8\pi P}\right)^{\frac{3}{2}}T >0,
\end{equation}
which is manifestly positive.\\
The computation of $c_V$ is more involved and can be carried out using the following relations:
\begin{align}\label{cp-cv}
   c_P-c_V&=TV\alpha_P^2k_T,\\\label{cv/cp}
    \frac{c_V}{c_P} &=k_T\beta_S,
\end{align}
where the isobaric thermal expansion coefficient, the isothermal bulk modulus and the adiabatic compressibility are given by, respectively
\begin{align}
    \alpha_P&=\frac{1}{V}\frac{\partial V}{\partial T}{\Bigg |}_{P},\\
    k_T&=-V\frac{\partial P}{\partial V}{\Bigg |}_{T},\\
    \beta_S&=-\frac{1}{V}\frac{\partial V}{\partial P}{\Bigg |}_{S}.
\end{align}
Eliminating $k_T$ from \eqref{cp-cv} and \eqref{cv/cp} and using \eqref{cp} yields
\begin{equation}\label{cv}
    c_V = \frac{S^2 \beta_S}{S \beta_S + TV \alpha_P^2}.
\end{equation}
Now, the volume \eqref{volume} can be express in terms of $P$ and $M$ and then, by virtue of the Smarr formula \eqref{smarr formula}, one can write
\begin{equation}\label{relaz V(T,P,S)}
    V=\frac{\sigma}{12}\left(\frac{3}{8\pi P}\right)^{\frac{3}{2}}\left(1+\frac{9\pi T^2}{2P}\right),
\end{equation}
which easily yields $\beta_S$ and $\alpha_P$. Substituting them in \eqref{cv} and using \eqref{relaz S(P,T)} results in
\begin{equation}
c_V=\frac{3 \sqrt{\frac{3 \pi }{2}} T \left(2 P-3 \pi  T^2\right)}{8 P^{3/2} \left(2 P+15 \pi  T^2\right)},
\end{equation}
which is independent on the parameter $\alpha$. To properly study the $T$ dependence of this function at fixed volume, pressure must be expressed in terms of $T$ and $V$ inverting \eqref{relaz V(T,P,S)}. This can be done numerically for different values of $T$ and $V$ yielding the results presented in Fig.~\ref{fig:cV(T)}.
We choose positive values of $T$ and, for every value of $V$, we checked that $V>V_{min}$.
We see that to any given value of the volume it corresponds a temperature $T^*$ below which the specific heat becomes positive. Since for $\alpha\geqslant l^2/(24\sqrt[3]{2})$ the black hole is always super-entropic, it is possible to simultaneously have $\mathcal{I}<1$ and $c_V>0$.\\
The situation is similar to the one observed in \cite{Cong2019}, where the behaviour of $c_V$ is opposite, being positive at large temperatures. However, the crucial difference is that here $c_P$ is always positive. This could suggest that super-entropic black holes can be in thermodynamic equilibrium since there are configurations in which both specific heats are positive.\\
However, for every value of $V$ there is always a temperature above which $c_V$ becomes negative.
This region could be excluded if there existed two separate branches of black hole solutions, as it happens for the SAdS case. However, such separation does not occur here as there is only one connected branch. As argued in \cite{Cong2019}, it is sufficient to have $c_V<0$ for at least some part of the branch to make the whole branch thermodynamically unstable.\\
Therefore, we conclude that the black hole solution we found satisfies the broader conjecture, proposed in \cite{Cong2019}, that black holes violating the reverse isoperimetric inequality are thermodynamically unstable.
\\We conclude this section with the case  $0<\alpha<l^2/(24\sqrt[3]{2})$. In this sector, the inequality is violated if $r_e>\bar{r}$, where $\bar{r}$ is the root of a fourth order
polynomial\footnote{Explicitly, $\bar{r}$ is the only positive real root of
\begin{align*}
    \mathcal{P}(r)&=
    110592 \alpha ^3 r^4+\left(13824 \alpha ^2 l^3-442368 \alpha ^3 l\right)r^3\\
    &+ \left(-9l^8+576 \alpha  l^6-34560 \alpha ^2 l^4+663552 \alpha ^3 l^2\right)r^2\\
    &+\left(8 l^9-576 \alpha  l^7+27648 \alpha ^2 l^5-442368 \alpha ^3 l^3\right)r
    -2 l^{10}\\
    &+144 \alpha  l^8-6912 \alpha ^2 l^6+110592 \alpha ^3 l^4.
\end{align*}}. It satisfies $\bar{r}>r_c$ and corresponds to a temperature $\bar{T}$ via \eqref{temperature}. In this case there will be super-entropicity only for $T>\bar{T}$. The value $\bar{T}$ can be either above or below the turning point $T^{*}$ where $c_V$ changes sign, depending on the specific value of $\alpha$. We found that for every $\alpha$ there is always a thermodynamic configuration, namely values of $V$ and $T$, such that $\bar{T}<T^{*}$. Therefore, in all cases there are super-entropic black holes with positive $c_V$. Since the behaviour of $c_V$ is independent on $\alpha$ the above discussion is valid also in this case.

\section{Conclusions}\label{conslusions}

In this paper we investigated the role of the Immirzi parameter, by promoting it to a dynamical scalar field in the framework of $f(R)$ gravity. In particular, we searched vacuum solutions with spherical symmetry, studying their properties both at a classical and semiclassical level.
The inclusion of the Holst term in the action requires dealing with the Palatini formulation of $f(R)$ theories, consistently with the role played by connections in standard formulation of LQG.
In addition to the Immirzi field $\gamma$, the resulting model features a scalar field $\phi$, that is the scalaron of $f(R)$ theories in the Jordan frame.
Both are responsible for a non vanishing torsion tensor, whose components are uniquely determined by the gradient of the scalar fields.
Exploiting this dependence, we derive a metric theory, dynamically equivalent, where the torsion degrees of freedom are reabsorbed in the non-standard kinetic terms of the scalar fields.
The structural equation governing the dynamics of $\phi$ turns out to be modified with respect to the standard case, and it acquires an additional term, which depends on the Immirzi field potential. This can sustain a non trivial profile for $\phi$, as opposed to ordinary Palatini $f(\mathcal{R})$ models, where it must boil down to a constant when the vacuum case is considered.
\\We then specialized to the vacuum spherically symmetric sector of the theory and, after selecting a Starobinsky-like $f(\mathcal{R})$ model and a potential for the Immirzi field, we found an analytical solution generalizing the one reported in \cite{Martinez2004}. It describes a locally asymptotic AdS black hole, whose event horizon has the peculiar topology of a genus $g\geqslant 2$ compact surface of constant negative curvature, with a horizon structure similar to \cite{PhysRevD.85.084035}. Beside the origin, there are curvature singularities at the roots of the conformal factor multiplying the metric tensor. Restricting the model parameter $\alpha$ to positive values, these are always hidden behind the black hole event horizon.
\\The black hole is endowed with secondary hair provided by the Immirzi field, which in turn implies a non trivial radial profile also for the scalaron via the modified structural equation. The scalar fields are regular everywhere on and outside the horizon, and they relax asymptotically, reducing to constant values $\gamma\rightarrow\gamma_0\equiv \sinh(\psi_0/\sqrt{3})$ and $\phi\rightarrow 1$. Therefore, in the asymptotic region the standard picture of LQG with a constant Immirzi parameter and a cosmological constant is recovered.
\\On the other hand, their effect becomes evident near the event horizon, especially regarding the thermodynamic properties of the black hole, which we investigated following the Euclidean path integral method and regularizing the action with the counter-terms method. The counter-terms suited for GR with a minimally coupled scalar field in asymptotically AdS spacetimes were derived in \cite{Gegenberg2003}. Here, we use instead a generalized version, suitable for the non-minimally coupled case at hand.
\\We also computed the black hole entropy applying Wald's method \cite{PhysRevD.48.R3427}, obtaining equivalent results. This is not a trivial outcome since in \cite{CHAKRABORTY2018432} it was shown that Wald's entropy formula is not affected by the presence of torsion which, however, the authors assumed to be non-dynamical, while here we deal with propagating torsional degrees of freedom.
\\Instead, if we had started adopting the standard view of a constant Immirzi parameter $\gamma(x)\equiv\gamma_0$, this would have implied a constant scalaron too, $\phi(x)\equiv\phi_0$, via \eqref{eq: struct eq Immirzi}. The resulting absence of torsion would have not affected the entropy computed via Wald’s formula which would have given an expression satisfying the usual area law ($\phi_0=1$), regardless of the specific expression of the metric functions.
\\The results emerging from this analysis allow discerning between an Immirzi parameter and an Immirzi field. We demonstrated, indeed, that the Immirzi field affect the entropy of the black hole via equation \eqref{entropy}, producing a modification with respect to the standard area law. On the other hand, a signature for the Immirzi field is expected to arise already in a classical scenario, namely in tidal forces experienced by infalling bodies. These are due to the geodesic deviation equation which is known to acquire corrections from non vanishing torsion components \cite{Speziale2018,Luz2017,Puetzfeld2018}, ultimately sourced by the Immirzi field via \eqref{eq: torsion components}. This draws attention to the definite mechanisms able to induce a dynamics for the Immirzi parameter, circumventing the unpleasant choice of promoting it to an additional degree of freedom by hand. In this sense, future investigations have to be devoted to the research of a unified kinematic setting, which could offer an elegant way for generating an Immirzi field, equipped with a potential term as well. We emphasize, moreover, that according a Palatini perspective, the inclusion of the Holst term in the Einstein action is not completely satisfactory, since in the presence of an Immirzi parameter GR is recovered only on half-shell, i.e. once Levi-Civita solution for the connection is obtained. It seems more reasonable, therefore, to enlarge our analysis to the Nieh-Yan term, which for an Immirzi parameter is genuinely topological even off-shell \cite{Nieh1982,Nieh2007}.
\\The presence of a negative cosmological constant allows to extend the thermodynamic phase space in line with \cite{Kubiznak2017}, including a pressure term in the first law of thermodynamics, together with its conjugate quantity, the thermodynamic volume.
The study of asymptotically AdS black holes in this extended thermodynamic phase space led to the proposition of a series of subsequent conjectures, each substituting the previous one whenever a new solution appeared to violate it.
The solution under study violates each of these conjectures except the last, which seems to be supported by the black hole analysed in this paper, although in a slightly different way with respect to the other two previously known examples (see discussion in section~\ref{BH thermo B}).\\
To see this, we first studied the thermodynamic volume and its relation with the horizon area encoded in the reverse isoperimetric inequality. We observed a violation of the inequality in almost all parameter space, implying the possibility of super-entropic black hole thermodynamic configurations.
This constitutes another example in contrast with the conjecture, initially proposed in \cite{Dolan2013}, that the thermodynamic volume satisfies the reverse isoperimetric inequality.
\\We note that violations of the RII never appeared in literature in solutions sharing the same properties of the one under consideration in this paper, namely the hyperbolic topology of the horizon and the presence of scalar hair surrounding it (super-entropic hairy black holes where found in \cite{Feng2017} for planar horizons).
In this regard we observe that the super-entropic behaviour of asymptotically AdS black holes seems to be a general feature independent on the specific peculiarities of each solution.
\\Moreover, being our solution characterized by a compact horizon, it is also in contrast with the broader conjecture proposed in \cite{PhysRevD.84.024037} that super-entropic black holes must have non compact horizons.
\\Finally, we investigated the thermodynamic stability of such super-entropic configurations, computing the specific heats at constant pressure $c_P$ and volume $c_V$. As it emerges from Fig.~\ref{fig:cV(T)}, a distinctive feature of the $c_V$ profile is the presence of Schottky-like peaks, which have already been suggested in \cite{Johnson2019} to be the evidence of finite energy windows at disposal for the excitation of underlying microscopic degrees of freedom.
Regarding their sign, $c_P$ turns out to be always positive but $c_V$ becomes negative at high enough temperatures, signalling a thermodynamic instability.
We conclude that the black hole solution studied in this paper supports the conjecture proposed in \cite{Cong2019,Johnson2020} that super-entropic black holes are thermodynamically unstable.

\acknowledgments

The work of F. B. is supported by the Fondazione Angelo della Riccia grant for the years 2020-2021. S. B. thanks Adolfo Cisterna for useful discussions.

\bibliography{references}{}

\end{document}